# Enhancing Conference Participation to Bridge the Diversity Gap

Laura Prichard[1]*, Cristina Oliveira[1], Alessandra Aloisi[1], Julia Roman-Duval[1], Svea Hernandez[1], Camilla Pacifici[1], Ivelina Momcheva[1], Space Telescope Science Institute Women in Astronomy Forum

[1]*Space Telescope Science Institute, 3700 San Martin Drive, Baltimore, MD 21218, USA*

**Abstract**

Conference attendance is fundamental for a successful career in astronomy. However, many factors limit such attendance in ways that can disproportionately affect women and minorities more. In this white paper, we present the results of a survey sent to 164 research staff at the Space Telescope Science Institute to determine what reasons motivate their attendance at science conferences and what aspects prevent researchers from attending them. The information collected through this survey was used to identify trends both in aggregate form and split by gender and if respondents had dependents. We propose a set of recommendations and best practices formulated by analyzing these trends. If consistently adopted, these recommendations will achieve greater diversity in astronomy through the broadening of conference participation.

## 1. Background and Motivation

Attending conferences is an essential part of career progression in astronomy. It is well known that one of the most critical components for obtaining a stable position in academia is being well-established and recognized within the scientific community. This makes conferences a valuable resource for sharing results, promoting work, networking, starting new collaborations, learning more about a field, and advancing research areas. However, there can be a large number of factors that limit researchers from attending conferences. Within a society that has systemic inequalities, it is likely that minority groups disproportionately face these limiting factors. This could result in conferences that are non-representative of the astronomical community's demographics worldwide, or that do not provide the same opportunities to all scientists unless additional practices to facilitate participation are put in place. These best practices, if consistently implemented, will ultimately result in greater diversity within the astronomical community as a whole.

Up to now, there has only been anecdotal evidence that there are key factors that affect people's ability to attend conferences; however, there is little research to document this in a statistically meaningful way. As an initiative by the Women in Astronomy Forum (WiAF) at the Space Telescope Science Institute (STScI), one of the largest astronomical facilities in the US, we decided to utilize the large number of research staff to conduct a survey on factors that affect a scientist's ability to attend conferences. STScI is large and relatively diverse in binary gender identity, number of dependents, and seniority. However, these demographics represent only a narrow portion of diversity and are not fully representative of the broader astronomical community. Ultimately, the "gap" we wish to bridge with this white paper extends beyond the experiences of staff at STScI and we hope the recommendations outlined here will benefit the astronomical community as a whole. In this paper, we present the survey design, respondent demographics, the key findings of the survey analysis, and outline recommendations for the community.

## 2. Design of the Survey

This anonymous survey was composed of 27 questions (given in Appendix A) and was designed to investigate three main elements: a) the motivation for researchers in astronomy to attend conferences; b) the factors limiting their ability to attend conferences; and c) identifying practices that have been or could be implemented to facilitate conference participation. In addition, the survey included optional questions to capture the demographics of the participants, including gender identity, race/ethnicity, age, years since PhD completion, type of position, relationship status, and number of dependents. A few of the questions had an exhaustive list of possible answers in order to bound the responses and facilitate the analysis, as well as a typable box for additional comments. A pilot survey was first conducted within the WiAF to refine the questions. It was then circulated more broadly to all the STScI research staff. Trends between the demographics and the three elements (i.e., motivation to attend conferences, limiting factors, and possible best practices) are explored in Section 4. The questions and their answer format are listed in Appendix A.

**3. Demographics of Participants**

The survey was circulated to the research staff at STScI on May 20, 2019 and was accepting responses for two weeks. The list of 164 research staff at STScI that the survey was sent to included researchers in the Association of Universities for Research in Astronomy (AURA), European Space Agency (ESA), ESA/AURA contracts, Canadian Space Agency (CSA), as well as postdoctoral and STScI fellows. By the end of the two-week period, 83 people had filled out the survey (51% of the research staff).

STScI research staff have different fractions of research time depending on their positions. For example, staff with positions equivalent to tenure track (including ESA affiliated Astronomers) spend 50% of their time on independent research (34% of the respondents). Scientist track and support scientists have 20% of their time for independent research with the possibility to "buy back" time up to 50% (40% of the respondents). CSA scientists have 30% of independent research time (3% of respondents), while postdocs and STScI fellows have 100% research time (13% of respondents). A small fraction of respondents (10%) are technical staff who buy back up to 50% of their time to perform independent research.

We provide the demographic information of the respondents (Table 1) to place the survey results in context. However, it is important to note that these demographics only represent about half of the research staff at STScI and should not be interpreted to represent STScI or the worldwide astronomical community as a whole. In Table 1, we show the demographic questions followed by the number of people responding to that question, as well as the breakdown of the responses. These questions were optional and have varying numbers of responses compared to the 83 total respondents. Caution should be used when interpreting the results of the survey due to small number statistics.

For the analysis discussed in the next sections, we use the term "women" to refer to respondents identifying as a "woman", and correspondingly for the term "men".

**Table 1** - Demographic information of respondents

| Demographic | Number of responses | Breakdown of responses |
|---|---|---|
| Gender identity | 73 | Man (68%), Woman (32%) |

| Years since completion of PhD* | 71 | Postdoc (1–6 years: 17%), Pre-tenure (7–11 years: 31%), Mid-career (12–18 years: 17%), Late-career (19+ years: 35%) |
| --- | --- | --- |
| Racial identity | 67 | White/Caucasian (95%), South Asian (1%), East Asian (1%), Two+ races - South Asian/White/Caucasian (1%) |
| Ethnic identity | 64 | Not Hispanic or Latinx (97%), Hispanic or Latinx (3%) |
| English as a first language | 73 | Yes (67%), No (33%) |
| Relationship status | 68 | Married/civil partnership (76%), In a relationship (7%), Single (7%), Engaged (4%), Divorced (3%), Domestic Partnership (1%) |
| Number of dependents | 73 | 0 (45%), 1 (18%), 2 (33%), 3 (4%) |

*Binned into coarser approximate career stage bins than the 1 year steps given in the survey (options given: 0–20+).

## 4. Results of the Survey

The results of the survey were analyzed in aggregate form and also by gender, number of dependents, and career stage, which are the demographic categories for which enough data was collected. When combining demographic information to explore trends in a quantitative way (e.g. disentangling career stage and number of dependents), we were limited by small numbers. Therefore, below we only present the most significant and clean results from the analysis. All respondents who answered the gender identity question (88% of all respondents) identified as either a man or woman and therefore the gender analysis only considers these binary genders. In the following subsections, we explore any observed trends between the demographics and the three elements introduced in Section 2: a) the motivation for researchers in Astronomy to attend conferences; b) the factors limiting their ability to attend conferences; and c) identifying practices that have been or could be implemented to facilitate conference participation.

### 4.1 Main Motivation to Attend Conferences

Most of the research staff at STScI attend conferences for several reasons, including presentation of new scientific results (94%), relevance of the conference topic to their work (86%), networking (78%), if they are invited to give a talk (74%), or to learn more about a given field (70%).

Results suggest that men tend to attend conferences more often if their abstract for a contributed talk is accepted, while women are more inclined to accept invitations for conferences, meetings and colloquia (Figure 1). The survey also shows that 22% (30%) of men (women) have not received an invitation for a talk in the past three years.

In general, men get invited to conferences more frequently than women: e.g., 78% of men received between 1 and 11 invitations over the last three years compared to only 52% of women. However, 16% of the 23 women (all either on Astronomer track or CSA staff but at mixed career stages) received at least 15 invitations or more. Compared to men, whose maximum number of invitations was 11, this is significantly higher. This trend may be the result of conference, meeting or colloquia organizers making efforts to increase the diversity of attendees or speakers with targeted invitations. However, it is evident that a large number of these invitations goes to only a small subset of women.

Based on the analysis above, we conclude that overall, women are invited less than men to conferences, meetings, and colloquia, but are more likely to accept invitations.

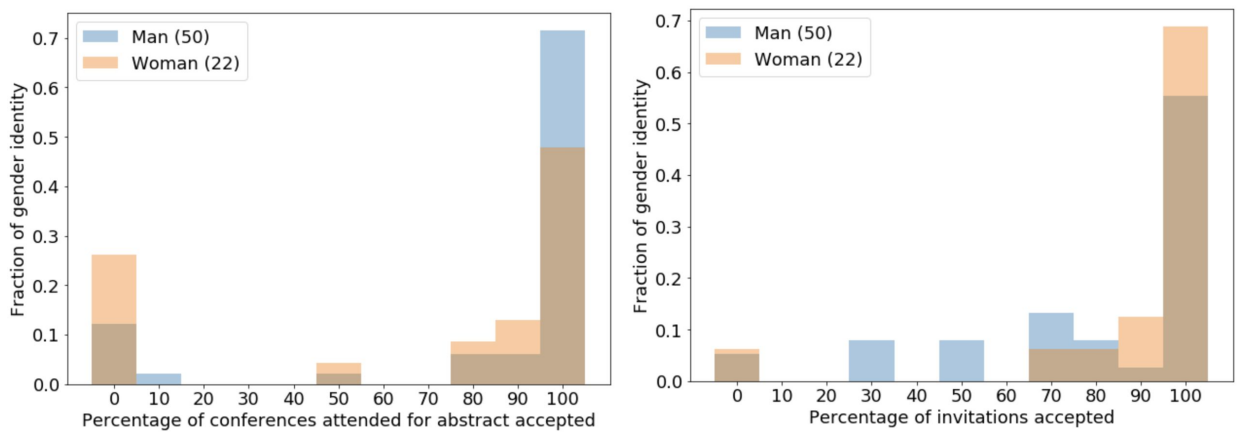

**Figure 1:** *Left:* Fraction of respondents by gender identity vs. percentage of conferences attended for abstracts accepted. *Right:* Fraction of respondents by gender identity vs. percentage of invitations to give a talk (in colloquia, meetings, or conferences) accepted.

### 4.2 Factors that Limit Conference Participation

In general, we found that the main reasons for not attending a conference or not submitting an abstract for a conference of interest include lack of time for conferences or other work commitments (49%), travel funds (42%), and too many conferences to attend (33%). When considering the main reasons for not attending conferences by gender, we did not find significant differences between men and women. A surprising 41% of the respondents did not accept one or more invitations to give a talk over the last 3 years. Our survey also showed that 55% of respondents would attend a conference if they were allocated a poster even if they had applied for a contributed talk. Overall, 75% of respondents had all their conference abstracts accepted (over the last 3 years), and 27% did not attend one or more conferences for which their abstract was accepted.

In order to further understand the factors that limit conference participation we looked at the number of conferences that people were interested in attending but ended up *not submitting an abstract for*. We also looked at the number of conferences for which the abstract was submitted but the respondents ended up *not attending the conference*.

The number of conferences that people were interested in attending over the last 3 years but decided not to submit an abstract for, was analyzed both in terms of gender identity and in terms of having no or any dependents (Figure 2). Overall, the distribution for men and women with no dependents (pale blue and orange, respectively) is the same and is peaked around an average of 2 conferences that respondents were interested in but did not submit an abstract for. When dependents are considered, the distribution of number of conferences people are interested in but end up not submitting an abstract for is different for men (dark blue) and women (red). Women with dependents tend to not submit abstracts at double the rate of men with dependents (6 vs. 3, respectively). Additionally, the difference between men with and without dependents is only 1 conference, while for women, that difference is more significant at 4 conferences.

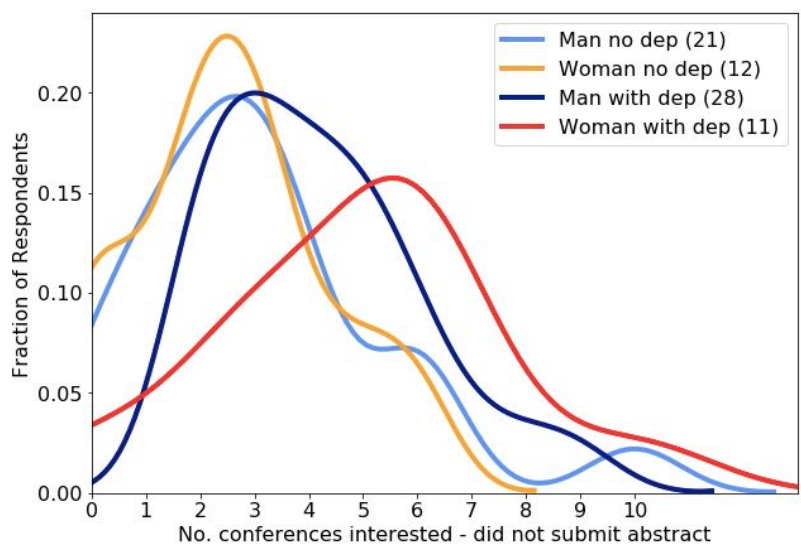

**Figure 2:** Fraction of respondents split by gender identity and by dependents or no dependents vs. the number of conferences that people were interested in attending but decided not to submit an abstract for, over the last three years.

When looking at the percentage of people that did not attend a conference even though they submitted an abstract (Figure 3), we found that ~20% of men that submitted an abstract did not attend the corresponding conferences. For women that percentage is much higher, with ~50% of the women not attending conferences for which they had submitted an abstract.

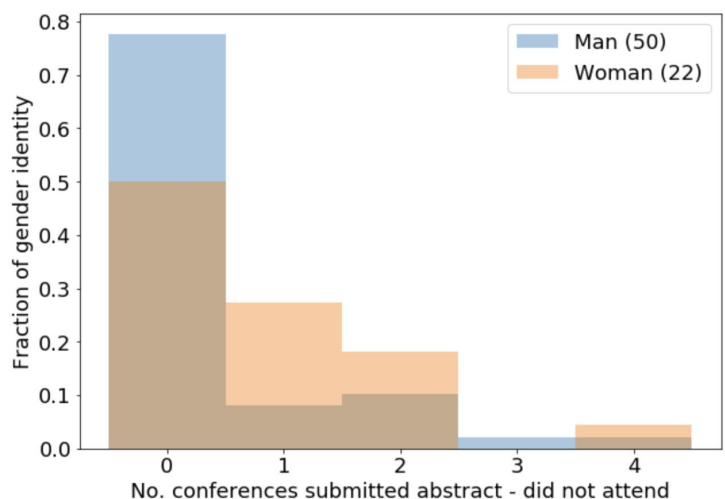

**Figure 3:** Fraction of respondents by gender identity vs. the number of conferences for which an abstract was submitted but the conference was not attended.

**4.3 Practices That Can Help with Conference Participation**

One of the main goals of this project was to provide recommendations on possible practices to enhance conference participation that could benefit the whole community. Respondents were asked to select all the practices/resources that would have enabled them to attend conferences, had they been offered. Overall, 34% of respondents said that no additional practices or resources were needed to enable their conference attendance. However, additional travel funding from grants/organizers and remote participation were selected each by 33% of the respondents as factors that would have enabled them to attend conferences.

When gender is considered, 40% of men stated that no additional resources would have enabled them to attend more conferences, compared to only 22% of the women. More women (43%, compared to 28% of men) said that remote participation would have enabled them to attend a conference. The survey also highlighted that gender balance and diverse representation of attendees was a factor that would have encouraged more women (22%, compared to 0% of men) to attend conferences.

**5. Recommendations and Conclusions**

The survey results presented above highlight how similar factors affect conference attendance differently for men and women and for people with and without dependents. In order to address this imbalance, we recommend the adoption of the following practices that we believe will benefit other minority groups within astronomy.

***Remote participation:*** allowing remote participation will preferentially facilitate participation of women in conferences. This resource may also improve participation of other underrepresented groups that may be more likely to be in smaller institutions with less funding for travel. Technology exists to enable remote participation and in fact there are a number of conferences that have already successfully offered this. The level of participation (one-way streaming, or two-way presenting and being involved in discussion remotely) will inevitably be at the discretion of the conference organizers and subject to local logistics. We recommend as an industry standard, to strive for two-way remote participation at every conference. In addition to the significant gains in inclusion, with over 300 conferences in astronomy listed in 2018 and 2019[1], travel as part of this profession notably contributes to climate change. Remote participation will help lessen the carbon footprint of astronomy and begin to address our part in this issue.

***Gender balance at conferences:*** having a diverse Science Organizing Committee and speaker list is important to achieve greater diversity of attendees. One area where some organizing committees are branching out is to select a diverse list of attendees, e.g., by pre-registration and final selection of participants through ad hoc software (e.g., `entrofy`, Huppenkothen et al. 2019) instead of a registration based on a "first-come first-served" basis. In addition, a balance of conference/colloquia speakers will provide more women with the same opportunities as men by being invited to more conferences. Our analysis emphasizes the importance of sending invitations to women to increase the chance their attendance and to make more effort to include junior women when trying to achieve greater diversity among speakers or attendees.

***Flexible funding for dependent care support:*** conference attendance is particularly critical for early and middle-career astronomers who are also more likely to have young dependents. As shown in Figure 2, women with dependents tend to apply to fewer conferences that they are interested in than men with dependents, or men and women without. While remote participation and having care facilities on conference premises might help, we believe that the best solution would be to allocate flexible funding for any additional dependent care costs incurred as a result of participating in a conference. People may require a range of options that can only be easily facilitated by providing the funding and flexibility for them to make arrangements that best suit their needs. These options might include funding for parents, dependents or caregivers to travel with the astronomer, local or onsite care, or more support at home while the astronomer is at the conference. We feel that this would be extremely beneficial to *all* astronomers with dependents. More established researchers could also benefit from this kind of flexible funding, as they are most often the caregivers of non self-sufficient elderly parents.

***Additional funding for travel:*** we recommend wherever possible, the allocation of institutional or conference funding to be used by participants to attend conferences. In particular, conference funding could be primarily used to target specific

---

[1] http://www.cadc-ccda.hia-iha.nrc-cnrc.gc.ca/en/meetings/getMeetings.html?year=2019&title=2019%20Meetings

underrepresented groups and improve overall diversity and inclusiveness at conferences.

We hope that these recommendations will stimulate conference organizers, funding agencies and institutions to broaden conference attendance through remote participation, increased diversity of organizers and invitees, and availability of funding to both support the speaker's travel and their dependent care. We point out that institutions such as the American Astronomical Society (AAS)[2] and ARC Centre of Excellence for All Sky Astrophysics in 3D (ASTRO 3D in Australia)[3] already have a flexible child/dependent care grants program in place. In addition, ASTRO 3D also requires that all Centre-run workshops or conferences have a 50% representation of women on the organizing committees, invited and contributed speakers and session chairs (see ASTRO 3D Diversity and Inclusion Action Plan).

We recommend a similar survey to be rolled out to a wider part of the astronomical community to capture responses and experiences that have not been covered here due to the specific composition of the STScI research staff and their limited viewpoints that may not be representative of the astronomical community as a whole.

**Appendix A - Survey questions**

Conference related questions:
1. Do you need to attend conferences to have a successful career in your field? [Yes/No]
2. What are some factors that make you want to attend a conference? (Please select all that apply, and specify any additional information or alternative responses in "Other") [Check-box multiple choice answers and text box provided]
3. How many science conferences that you were interested in have you *decided not to submit an abstract for* in the past three years? [Numerical]
4. How many science conferences *did you not attend* in the past three years even though you submitted an abstract? [Numerical]
5. If your answers to question 3 or 4 are more than zero: What were the factors preventing you from participating in those events? (Please select all that apply, and specify any additional information or alternative responses in "Other") [Check-box multiple choice answers and text box provided]
6. If your answers to question 3 or 4 are more than zero: If any of the following resources/practices had been offered, would you have attended those events? [Check-box multiple choice answers and text box provided]
7. Are there already resources/practices offered that facilitated your attendance at previous conferences? [Yes/No/Text box]
8. How many invitations to speak (at a conference, meeting, or colloquium) did you receive in the last three years? [Numeric]

---

[2] https://aas.org/grants-and-prizes/childdependent-care-grants
[3] https://astro3d.org.au/diversity-and-inclusion/

9.  Out of those invitations, how many did you accept and went on to give an invited talk? [Numeric]
10. What percentage of your submitted conference abstracts have been declined in the past three years? (Please give your best approximation if you cannot recall exactly) [Percentage]
11. Out of those accepted, what percentage did you attend? [Percentage]
12. Would you attend a conference if you were allocated a poster instead of a contributed talk? [Yes/No/Text box]

Optional demographic questions:
13. Position [At STScI, options and text box provided]
14. Average research fraction over the last three years (nominal + buy-back) [At STScI, percentage]
15. Years since completion of PhD [Numeric]
16. What is your gender identity? [Check-box multiple choice answers and text box provided]
17. Do you identify as transgender? [Yes/No/Prefer not to say]
18. What is your age? [Age bins provided]
19. What race do you identify as? [Check-box multiple choice answers and text box provided]
20. What ethnicity do you identify as? [Check-box multiple choice answers and text box provided]
21. Is English your first language? [Yes/No/Prefer not to say]
22. If you answered "No" to the previous question, do you consider yourself fluent in English? [Yes/No/Prefer not to say]
23. What is your relationship status? [Check-box multiple choice answers and text box provided]
24. How many dependents do you have? [Numeric]
25. What is the age of your dependents? [Multiple choice age bins provided]
26. Do or did (last three years) you have pets that require regular care? [Yes/No]
27. Please give any other comments relevant to this survey that you would like to share [Text box provided]

**Acknowledgements**

We thank the research staff at STScI for their willingness to participate in the survey and provide genuine responses that allowed the analysis presented in this paper to be performed. The authors would also like to thank Antonella Nota for fully encouraging and supporting this effort and for valuable contributions to the discussion. We would also like to thank Peter Zeidler who provided feedback on the white paper.